\documentclass[hyper]{JHEP} 

\usepackage{epsfig}

\newcommand\fverb{\setbox\pippobox=\hbox\bgroup\verb}

\newcommand\fverbdo{\egroup\medskip\noindent%

            \fbox{\unhbox\pippobox}\ }

\newcommand\fverbit{\egroup\item[\fbox{\unhbox\pippobox}]}

\newbox\pippobox

\title{Non-Relativistic D-brane from T-duality Along
Null Direction}
\author{J. Kluso\v{n}\\
Department of
Theoretical Physics and Astrophysics\\
Faculty of Science, Masaryk University\\
Kotl\'{a}\v{r}sk\'{a} 2, 611 37, Brno\\
Czech Republic\\
E-mail: \email{klu@physics.muni.cz}} \preprint{}

 \abstract{We construct new type of non-relativistic
D-branes which are defined with the help of T-duality
along null direction. We find Lagrangian and Hamiltonian
formulation of these D-branes and study their properties
under T-duality transformations.}

\def\hv{\hat{v}}
\def\he{\hat{e}}

\def\hgamma{\hat{\gamma}}

\def\balpha{\bar{\alpha}}
\def\bbeta{\bar{\beta}}

\def\ty{\tilde{y}}

\def\bA{\mathbf{A}}

\def\bF{\mathbf{F}}

\def\be{\begin{equation}}

\def\halpha{\hat{\alpha}}
\def\hbeta{\hat{\beta}}
\def\ee{\end{equation}}

\def\bea{\begin{eqnarray}}
\def\bh{\bar{h}}
\def\hi{\hat{i}}
\def\hj{\hat{j}}
\def\eea{\end{eqnarray}}

\def\bmu{\bar{\mu}}
\def\bnu{\bar{\nu}}

\def\mH{\mathcal{H}}

\def\hbA{\hat{\bA}}

\def\hbA{\hat{\bA}}

\newcommand{\mG}{\mathcal{G}}

\newcommand{\hk}{\hat{k}}

\def \bAi{\left(\mathbf{A}^{-1}\right)}

\def \bA{\mathbf{A}}

\newcommand{\ba}{\mathbf{a}}

\newcommand{\mL}{\mathcal{L}}

\def\pb #1{\left\{#1\right\}}

\begin{document}
\section{Introduction and Summary}
There was a renewed interest in the Newton Cartan (NC) \cite{Cartan:1923zea} geometry and its
torsionful generalization (TNC) (see for example
\cite{Bergshoeff:2017dqq,Bergshoeff:2014uea}) since NC geometry is very useful tool for the study of non-relativistic field theories, string theories and geometry. More precisely, torsionful NC geometry, which is characterized by non-exact clock form was observed as the boundary geometry
in the context of Lifschitz geometry
\cite{Christensen:2013lma,Christensen:2013rfa,Hartong:2014pma}. Further, NC geometry can be also
applied for formulation of non-relativistic string theories
\cite{Andringa:2012uz,Harmark:2017rpg,Kluson:2018egd,Kluson:2018uss,Bergshoeff:2018yvt,
Harmark:2018cdl,Gomis:2019zyu,Gallegos:2019icg,Harmark:2019upf,Grosvenor:2017dfs,Kluson:2019uza,Kluson:2019ifd}
 that could be UV completition of non-relativistic gravity. All there works are based on two seminal papers \cite{Gomis:2000bd,Danielsson:2000gi} where independently an action for non-relativistic string in flat background was proposed. It is interesting that there are currently two versions of non-relativistic strings. The first one is based on the null reduction that leads to Nambu-Goto action
 \cite{Harmark:2017rpg} and Polyakov like action \cite{ Harmark:2018cdl}
 on torsional NC geometry with extra periodic dimension. There is an alternative definition of non-relativistic strings that is based on
large $c$ expansion of action for string in stringy NC geometry
\cite{Andringa:2012uz,Bergshoeff:2018yvt}. Very recently the relation between these two theories was found in  interesting paper \cite{Harmark:2019upf}.

In this paper we focus on the first approach that defines non-relativistic string theories based on the null reduction. More precisely, as was shown in \cite{Kluson:2018egd}, the non-relativistic string arises from T-duality along  null direction. Since it is well known that string theory contain another extended objects in their spectrum, as for example D-branes \cite{Polchinski:1995mt}, one can ask the question whether it is possible to define non-relativistic Dp-branes exactly in the same way. More precisely, we consider Dp-brane in the background with null isometry and we presume that Dp-brane wraps this  direction. Then applying T-duality along this direction  we derive non-relativistic D(p-1)-brane in T-dual geometry that is localized along T-dual  dimension
\footnote{For earlier work, see 
	\cite{Mukhi:2002ck}.}. We also find its Hamiltonian formulation when we identify diffeomorphism and Hamiltonian constraints.

As the next step in our analysis we will study properties of this non-relativistic D(p-1)-brane when it wraps another compact direction. We find that T-dual object is D(p-2)-brane localized along this compact direction on certain condition that the background field has to obey.

We also show an alternative way how to define Hamiltonian for non-relativistic D(p-1)-brane in torsional NC geometry which is based on the canonical description of T-duality
\footnote{For review, see for example
\cite{Simon:2011rw}.}. We firstly find Hamiltonian for relativistic Dp-brane in general background and we show how T-duality can be defined in the canonical formalism. More precisely, in order to perform T-duality along compact dimension we have to introduce gauge fixing function that relates embedding coordinate with the world-volume one. This gauge fixing function together with one  spatial diffeomorphism constraint are second class constraints that can be explicitly solved. It turns out that it is crucial that there is a gauge field propagating on the world-volume of
Dp-brane for solving  spatial diffeomorphism  constraint since in the absence of the gauge world-volume field it would not be possible to introduce coordinate that localises D(p-1)-brane in dual dimension.  We show that the resulting Hamiltonian describes D(p-1)-brane in T-dual background with the background metric, NSNS two form and dilaton that  are given by famous Buscher's rules \cite{Buscher:1987sk,Buscher:1987qj}. Then we apply the same procedure to the  case of Dp-brane in the background with
null isometry  and we derive Hamiltonian for D(p-1)-brane in NC background that coincides with the Hamiltonian that was derived independently in the section devoted
to Lagrangian formulation of non-relativistic D-brane. We also analyze an explicit example of non-relativistic D1-brane with the fixed electric flux which corresponds to the number of fundamental strings bounded to D1-brane. We determine corresponding Lagrangian density.

Let us outline our results and suggest possible extension of this work. We define new non-relativistic D(p-1)-branes in TNC geometry and we study their properties under T-duality transformations. We find that non-relativistic D-brane transforms covariantly under T-duality transformation. We also discuss an explicit example of non-relativistic
D1-brane that has similar structure as corresponding non-relativistic string which is
clearly seen in the Hamiltonian formulation of both theories.

This paper can be extended in many directions. For example, it would be nice to generalize this construction to more general background with non-zero NSNS two forms and Ramond-Ramond fields. It would be also nice to study super D-branes along this direction. Finally, it would be nice to analyze non-BPS Dp-branes in the similar way and study tachyon condensation in this context. We hope to return to some of these problems in future.

This paper is organized as follows. In the next section (\ref{second}) we find non-relativistic D-brane in the Lagrangian formalism. Then in section (\ref{third}) we find its Hamiltonian form. In section (\ref{fourth}) we study T-duality of non-relativistic D-brane. In section (\ref{fifth}) we review canonical description of T-duality of D-brane and finally in section (\ref{sixth}) we apply this analysis to the definition of the canonical form of non-relativistic D-brane in TNC background.

\section{Lagrangian Formulation of Non-Relativistic D-brane}\label{second}
In this section we formulate non-relativistic D-brane in the Lagrangian formalism.
The starting point is an action for Dp-brane in general background that has the form
\begin{equation}
S=-T_p\int d^{p+1}\xi e^{-\phi}\sqrt{-\det\bA_{\alpha\beta}} \ ,  \quad
\bA_{\alpha\beta}=g_{\alpha\beta}+b_{\alpha\beta}+l_s^2 F_{\alpha\beta} \ ,
\end{equation}
where $\xi^\alpha, \alpha,\beta=0,1,\dots,p$   label world-volume of Dp-brane.
Further, $g_{\alpha\beta},b_{\alpha\beta}$ are pullbacks of the background fields
$G_{MN},B_{MN}$
\begin{equation}
g_{\alpha\beta}=G_{MN}\partial_\alpha x^M\partial_\beta x^N \ , \quad
b_{\alpha\beta}=B_{MN}\partial_{\alpha}x^M\partial_\beta x^N \ ,
\end{equation}
where $x^M(\xi)$ parameterize embedding of Dp-brane in target space-time.
$F_{\alpha\beta}=\partial_\alpha A_\beta-\partial_\beta A_\alpha$ is field strength
of the world-volume gauge field $A_\alpha$ which is crucial for the definition of
Dp-brane. Finally, $T_p=\frac{1}{l_s^{p+1}}$ is Dp-brane tension and $l_s$ is string
length.

Our goal is to define non-relativistic D(p-1)-brane  as T-dual of relativistic Dp-brane
in the background with null isometry. This approach is based on seminal paper
\cite{Harmark:2017rpg} where an action for non-relativistic string was proposed.
Then it was shown in \cite{Kluson:2018uss} that such a string corresponds to the string that is T-dual to the string in the null background. Note that the null background is defined by following background metric
\cite{Harmark:2017rpg}
\begin{equation}\label{backmet}
ds^2=g_{MN}dx^M dx^N=2\tau (du-m)+h_{\mu\nu}dx^\mu dx^\nu \ , \quad
\tau=\tau_\mu dx^\mu \ , \quad  m=m_\mu dx^\mu \ ,
\end{equation}
where $\det h_{\mu\nu}=0$. We implicitly consider superstring theory so that  $M,N=0,1,\dots,9$ while $\mu,\nu=0,1,\dots,8$. The inverse metric has the form
\begin{eqnarray}\label{backmetin}
& &G^{uu}=2\Phi \ , \quad G^{u\mu}=-\hv^\mu ,  \quad G^{\mu\nu}=h^{\mu\nu} \ ,
\nonumber \\
& & \Phi=-m_\mu v^\mu+\frac{1}{2}m_\mu h^{\mu\nu}m_\nu \ , \quad
\hv^\mu=v^\mu-h^{\mu\nu}m_\nu \ .  \nonumber \\
\end{eqnarray}
Note that the inverse metric $h_{\mu\nu}$ and $v^\mu$ obey the relation
\begin{equation}
h_{\mu\nu}h^{\nu\rho}-\tau_\mu v^{\rho}=\delta_\mu^\rho \ .
\end{equation}
Now we presume that Dp-brane is extended along $u-$direction where
the background posses an isometry $u\rightarrow u+\epsilon$. Then it is natural to perform T-duality along this direction so that we will have D(p-1)-brane localized along dual direction. More precisely, let Dp-brane wraps $u-$direction so that we have static
gauge along $\xi^p$ coordinate
\begin{equation}
u=\xi^p
\end{equation}
and all fields do not depend on $\xi^p$. Let $\halpha,\hbeta$ denote
remaining coordinates $\halpha=0,1,\dots,p-1$. Then the matrix $\bA_{\alpha\beta}$
has the form
\begin{eqnarray}
& &\bA_{\halpha\hbeta}=\hat{\bA}_{\halpha\hbeta}=\bh_{\mu\nu}
\partial_{\halpha}x^\mu\partial_{\hbeta}x^\nu+l_s^2 F_{\halpha\hbeta} \ , \nonumber \\
& &\bA_{\halpha p}= \partial_{\halpha}x^\mu \tau_\mu+l_s^2\partial_{\halpha}A_p \ , \nonumber \\
& &\bA_{p\hbeta}=\tau_\mu\partial_{\hbeta}x^\mu -l_s^2 \partial_{\hbeta}A_p
\ , \quad  \bA_{pp}=0 \ .  \nonumber \\
\end{eqnarray}
It is natural to presume that $\hbA_{\halpha\hbeta}$ is non-singular matrix with inverse
$\hbA^{\hbeta\halpha}$ so that $\hbA_{\halpha\hbeta}\hbA^{\hbeta\hgamma}=\delta_{\halpha}^{\hgamma}$. Then the determinant $\det\bA_{\alpha\beta}$ is equal to
\begin{equation}
\det\bA_{\alpha\beta}=\left|\begin{array}{cc}
\hat{\bA}_{\halpha\hbeta} & \bA_{\halpha p} \\
\bA_{p\hbeta} & 0 \\ \end{array}\right|=
\left|\begin{array}{cc}
\hat{\bA}_{\halpha\hbeta} & \bA_{\halpha p} \\
0 & -\bA_{p\halpha}\hat{\bA}^{\halpha\hbeta}
\bA_{\hbeta p} \\ \end{array}\right|
\end{equation}
and hence an action has the form
\begin{eqnarray}\label{p1brane}
& &S=-T_p\int du \int d^{p-1}\xi e^{-\phi}
\sqrt{\det\hat{\bA}_{\halpha\hbeta}}\sqrt{
	\bA_{p\halpha}\hat{\bA}^{\halpha\hbeta}
	\bA_{\hbeta p}}=\nonumber \\
& &=-T_{p-1}\int d^{p-1}\xi e^{-\phi' }
\sqrt{\det\hat{\bA}_{\halpha\hbeta}}
\sqrt{(\tau_{\halpha}-\partial_{\halpha}\eta)\hat{\bA}^{\halpha\hbeta}
	(\tau_{\hbeta}+\partial_{\hbeta}\eta)} \ ,
\nonumber \\
\end{eqnarray}
where we defined $\eta$ as $\eta=l_s^2 A_p$. We further included integration
over $u$ into definition of T-dual tension $T_{p-1}$ together to T-dual
dilaton as
\begin{equation}\label{defphi}
e^{-\phi}T_{p}\int du=e^{-\phi'}T_{p-1} \ .
\end{equation}
The action (\ref{p1brane}) is our proposal for non-relativistic D(p-1)-brane. Clearly it is invariant under world-volume diffeomorphism and it is also manifestly covariant. It is also instructive to give an explicit example of D1-brane.

In this case the  matrix inverse $\hat{\bA}^{\halpha\hbeta}$ has following  form
\begin{equation}
\hat{\bA}^{\halpha\hbeta}=\frac{1}{\det\hat{\bA}_{\halpha\hbeta}}
\left(\begin{array}{cc}
\hat{\bA}_{\sigma\sigma} & -\hat{\bA}_{\tau\sigma} \\
-\hat{\bA}_{\sigma\tau} & \hat{\bA}_{\sigma\sigma} \\
\end{array}\right)
\end{equation}
so that the action has the form
\begin{eqnarray}
S=-T_1 \int d^2\xi
e^{-\phi}(\partial_0\eta^2 \bh_{11}-\tau_0^2\bh_{11}+\partial_1\eta^2\bh_{00}
-\tau_1^2\bh_{00}-\nonumber \\
-2\partial_0\eta \partial_1\eta \bh_{01}+2\tau_0\tau_1\bh_{01}+2l_s^2
\tau_0\partial_1\eta F_{01}-2l_s^2\partial_0\eta
\tau_1 F_{01})^{1/2} \nonumber \\
\end{eqnarray}
introducing $\epsilon^{\halpha\hbeta}=-\epsilon^{\hbeta\halpha} \ , \epsilon^{01}=1$
we can rewrite this action into more elegant form
\begin{equation}
S=-T_1\int d^2\xi e^{-\phi} \sqrt{\partial_{\halpha}\eta\partial_{\hbeta}\eta
	\epsilon^{\halpha\halpha'}
	\epsilon^{\hbeta\hbeta'}
	\bh_{\halpha'\hbeta'}-
	\tau_{\halpha}\tau_{\hbeta}
	\epsilon^{\halpha\halpha'}
	\epsilon^{\hbeta\hbeta'}
	\bh_{\halpha'\hbeta'}+2l_s^2
	\tau_{\halpha}\partial_{\hbeta}\eta
	\epsilon^{\halpha\halpha'}\epsilon^{\hbeta\hbeta'}F_{\halpha'
		\hbeta'}}
\ . \end{equation}
We see that the expression under square root has similar structure as non-relativistic string action
that was found in \cite{Harmark:2017rpg} which is nice consistency check of our proposal.
\section{Hamiltonian Formalism for Non-Relativistic D-brane}\label{third}
In this section we find Hamiltonian form of the non-relativistic D-brane
given by the action (\ref{p1brane}).
It turns out that it is much simpler to consider gauge fixed matrix $\bA_{\alpha\beta}$ with its inverse that we denote as   $H^{\alpha\beta}$. Then it is easy to find following conjugate momenta
\begin{eqnarray}\label{defmomenta}
& &p_{\mu}=\frac{\partial \mL}{\partial (\partial_0 x^{\mu})}=T_{p-1}e^{-\phi}\bh_{\mu\nu}\partial_{\halpha}x^\nu H^{\halpha 0}_S
\sqrt{-\det\bA}+T_{p-1}e^{-\phi}\tau_\mu  H^{p 0}_S\sqrt{-\det\bA} \ , \nonumber \\
& &p_\eta=\frac{\partial \mL}{\partial (\partial_0 \eta)}=T_{p-1}e^{-\phi}H^{p0}_A\sqrt{-\det\bA} \ , \nonumber \\
& & \pi^{\hi}=\frac{\partial \mL}{\partial(\partial_0 A_{\hi})}l_s^2T_{p-1}e^{-\phi}H^{\hi 0}_A\sqrt{-\det\bA} \ , \quad \pi^0=\frac{\partial
\mL}{\partial(\partial_0 A_0)}\approx 0  \ ,
\nonumber \\
\end{eqnarray}
where
\begin{equation}
H^{\halpha\hbeta}_S=\frac{1}{2}(H^{\halpha\hbeta}+H^{\hbeta\halpha}) \ ,
\quad
H^{\halpha\hbeta}_A=\frac{1}{2}(H^{\halpha\hbeta}-H^{\hbeta\halpha}) \ .
\end{equation}
With the help of these results we obtain that the bare Hamiltonian density is equal to
\begin{eqnarray}\label{Hbare}
\mH_B=p_\mu\partial_0 x^\mu+p_\eta \partial_0 \eta+\pi^{\hi}\partial_0 A_{\hi}
-\mL
=\partial_{\hi}A_0
\pi^{\hi} \
\nonumber \\
\end{eqnarray}
and also we can identify $p-1$ primary constraints
\begin{equation}
\mH_{\hi}\equiv \partial_{\hi}x^\mu p_\mu+\partial_{\hi}\eta p_\eta+F_{\hi \hj}\pi^{\hj} \approx 0 \ .
\end{equation}
Further, (\ref{Hbare}) implies that there should exist one Hamiltonian constraint which
also follows from the fact that action for non-relativistic D(p-1)-brane is manifestly diffeomorphism invariant. In order to find such a constraint let us calculate following expressions
\begin{eqnarray}
& &p_\mu h^{\mu\nu}p_\nu=T_{p-1}^2 e^{-2\phi}H^{0\halpha}_S( \bh_{\halpha\hbeta}+
2\tau_{\halpha} \Phi \tau_{\hbeta})
H^{\hbeta 0}_S(\sqrt{-\det\bA})^2 \ , \nonumber \\
& & l_s^{-4}\pi^{\hi}\partial_{\hi}x^\mu \bh_{\mu\nu}
\partial_{\hj}x^\nu\pi^{\hj}=T_{p-1}^2 e^{-2\phi}
H^{\halpha 0}_A \bh_{\halpha\hbeta}H^{\hbeta 0}_A(\sqrt{-\det\bA})^2 \ ,
\nonumber \\
& &2l_s^{-4}\partial_{\hi}\eta \pi^{\hi}\Phi
\partial_{\hj}\eta\pi^{\hj}=2T_{p-1}^2 e^{-2\phi}\Phi H^{0\halpha}_A \partial_{\halpha}
\eta \partial_{\hbeta}\eta H^{\hbeta 0}_A(\sqrt{-\det \bA})^2 \ ,
\nonumber \\
& &-2l_s^{-2}\partial_{\hi}\eta \pi^{\hi}\hv^\mu p_\mu=
\nonumber \\
& & =-4T_{p-1}^2e^{-2\phi}\partial_{\halpha}\eta H^{\halpha 0}_A\Phi \tau_{\hbeta}H^{\hbeta 0}_S
(\sqrt{-\det\bA})^2+2
T_{p-1}^2 e^{-2\phi}\partial_{\halpha}\eta H^{\halpha 0}_A
H_S^{p0}(\sqrt{-\det \bA})^2 \ , \nonumber \\
& & 2l_s^{-2}p_\eta \tau_{\hi}\pi^{\hi}=2T_{p-1}^2 e^{-2\phi}\tau_{\halpha}H^{\halpha 0}_A
H^{p0}_A(\sqrt{-\det \bA})^2 \ , \nonumber \\
\end{eqnarray}
where we used
\begin{equation}
\bh_{\mu\nu}h^{\nu\delta}\bh_{\delta\omega}=
\bh_{\mu\omega}+2\tau_\mu \Phi\tau_\omega \ , \quad \hv^\mu \bh_{\mu\rho}=
2\tau_\rho \Phi \ .
\end{equation}
If we take all terms given above together we firstly find that expressions
proportional to $\Phi$ cancel each other.
%
In order to deal with remaining terms we use following expressions
\begin{eqnarray}
& & H^{\halpha\hbeta}_SF_{\hbeta\hgamma}-H_S^{\halpha p}\partial_{\hgamma}\eta+
H^{\halpha\hbeta}_A\bh_{\hbeta\hgamma}+H_A^{\halpha p}\tau_{\hgamma}=0 \ , \nonumber \\
& &\bh_{\halpha\hbeta}H_S^{\hbeta \hgamma}+
\tau_{\halpha}H^{p\hgamma}_S+l_s F_{\halpha\hbeta}H_A^{\hbeta\hgamma}+
\partial_{\halpha}\eta H_A^{p\hgamma}=\delta^{\hgamma}_{\halpha}\nonumber \\
& &H^{p\halpha}_S\partial_{\halpha}\eta+H^{p\halpha}_A\tau_{\halpha}=0 \ , \quad  
H^{\hgamma \halpha}_S\tau_{\halpha}+ H^{\hgamma	\halpha}_A\partial_{\halpha}\eta=0 \ , \nonumber \\
\end{eqnarray}
since $\bA_{pp}=0$.
Using these terms together we derive after some calculations following primary
Hamiltonian
constraint
\begin{eqnarray}
& &\mH_\tau=p_\mu h^{\mu\nu}p_\nu+l_s^{-4}\pi^{\hi}\partial_{\hi}x^\mu
\bh_{\mu\nu}\partial_{\hj}x^\nu \pi^{\hj}-2l_s^{-2}\partial_{\hi}
\eta \pi^{\hi}\hv^\mu p_\mu+2l_s^{-2}p_\eta\tau_{\hi}\pi^{\hi}+2l_s^{-2}
\partial_{\hi}\eta \pi^{\hi}\Phi \partial_{\hj}\eta\pi^{\hj}+
\nonumber \\
& &+T_{p-1}^2 e^{-2\phi}\det\ba_{\hi\hj}
(\partial_{\hi}\eta-\tau_{\hi})\ba^{\hi \hj}
(\partial_{\hj}\eta+\tau_{\hj})\approx 0  \ ,
\nonumber \\
\end{eqnarray}
where $\ba^{\hi \hj}$ is matrix inverse to the matrix  $\ba_{\hi \hj}=\bh_{\hi \hj}+l_s^2 \bF_{\hi \hj}$.

In this section we found Hamiltonian for non-relativistic D(p-1)-brane. The crucial fact is in the presence of the momentum $\pi^{\hi}$ that multiplies $p_\eta$ and $\hv^{\mu}p_\mu$. In case when $\pi^{\hi}$ were zero we would get Hamiltonian constraint $\mH_\tau=p_\mu h^{\mu\nu}p_\nu+
T_{p-1}^2 e^{-2\phi}\det \ba_{\hi\hj}(\partial_{\hi}\eta-\tau_{\hi})
\ba^{\hi\hj}(\partial_{\hj}\eta+\tau_{\hj})$. This is positive quantity that
cannot be equal to zero except exceptional cases as $p_\mu=0 \ ,\partial_{\hi}\eta=\tau_{\hi}$ or $\partial_{\hi}=-\tau_{\hi}$. Further, there is no term proportional to momentum conjugate to $p_\eta$. These reasons suggest
that non-relativistic D(p-1)-brane is well defined in case of non-zero electric flux only.
\section{T-duality along Longitudinal Direction}\label{fourth}
An important property of relativistic Dp-brane is its behaviour under T-duality when Dp-brane maps covariantly to D(p-1)-brane where the background fields are determined by
Buscher's rules \cite{Buscher:1987sk,Buscher:1987sk}. It is certainly interesting question to analyse similar situation in case
of non-relativistic D(p-1)-brane. To do this let us presume that D(p-1)-brane
wraps additional compact direction, say $y=x^8$. The fact that D(p-1)-brane wraps
this direction means that we impose the static gauge along this direction
\begin{equation}
 y=\xi^{p-1} \ .
\end{equation}
Further, all world-volume fields depend on $\xi^{\balpha} \ , \bbeta,\balpha=0,1,\dots,p-2$.
Then the matrix  $\bA_{\alpha\beta}$ has the form
\begin{eqnarray}
\bA_{\alpha\beta}=\left(\begin{array}{ccc}
\bA_{\balpha\bbeta} & \bA_{\balpha (p-1)} & \bA_{\balpha p} \\
\bA_{(p-1)\bbeta} & \bA_{(p-1)(p-1)} & \bA_{(p-1)p} \\
\bA_{p\bbeta}& \bA_{p(p-1)} & 0 \\ \end{array}\right) \ .
\nonumber \\
\end{eqnarray}
Now using properties of the determinant of the matrix $\bA_{\alpha\beta}$
we obtain
\begin{eqnarray}
& &\det \bA_{\alpha\beta}=
\left|\begin{array}{ccc}
\bA_{\balpha\bbeta} & \bA_{\balpha p} & \bA_{\balpha (p-1)} \\
\bA_{p\bbeta} & 0 &  \bA_{p(p-1)} \\
\bA_{(p-1)\bbeta} & \bA_{(p-1)p} & \bA_{(p-1)(p-1)} \\ \end{array}\right|=
\nonumber \\
& &=\left|\begin{array}{ccc}\
\bA_{\balpha\bbeta}-\bA_{\balpha (p-1)}\frac{1}{\bA_{(p-1)(p-1)}}
\bA_{(p-1)\bbeta} & \bA_{\balpha p}-\frac{\bA_{(p-1)p}}{\bA_{(p-1)(p-1)}}
\bA_{\balpha (p-1)} & 0 \\
\bA_{p\bbeta}-\bA_{(p-1)\bbeta}\frac{1}{\bA_{(p-1)(p-1)}}\bA_{p(p-1)} & -\frac{1}{\bA_{(p-1)(p-1)}}
\bA_{(p-1)p}\bA_{p(p-1)} & 0 \\
\bA_{(p-1)\bbeta} & \bA_{(p-1)p} & \bA_{(p-1)(p-1)} \\
\end{array}\right|=
\nonumber \\
& &=\bA_{(p-1)(p-1)}
\left|\begin{array}{cc}\
\bA_{\balpha\bbeta}-\bA_{\balpha (p-1)}\frac{1}{\bA_{(p-1)(p-1)}}
\bA_{(p-1)\bbeta} & \bA_{\balpha p}-\frac{\bA_{(p-1)p}}{\bA_{(p-1)(p-1)}}
\bA_{\balpha (p-1)}  \\
\bA_{p\bbeta}-\bA_{(p-1)\bbeta}\frac{1}{\bA_{(p-1)(p-1)}}\bA_{p(p-1)} & -\frac{1}{\bA_{(p-1)(p-1)}}
\bA_{(p-1)p}\bA_{p(p-1)}   \\
\end{array}\right| \ ,
\nonumber \\
\end{eqnarray}
where
\begin{eqnarray}
& &\bA_{(p-1)(p-1)}=\bh_{yy} \ ,  \quad
\bA_{(p-1)p}=\tau_y \ , \quad \bA_{p(p-1)}=\tau_y \ , \nonumber \\
& &\bA_{\balpha p}=\tau_{\bmu}\partial_{\balpha}x^{\bmu}+\partial_{\balpha}\eta \ , \quad
\bA_{p\bbeta}=\tau_{\bmu}\partial_{\bbeta}x^{\bmu}-\partial_{\bbeta}\eta \ , \nonumber \\
& & \bA_{\balpha (p-1)}=\partial_{\balpha}x^{\bmu}\bh_{\bmu y}+\partial_{\balpha}\ty \ ,
\quad
\bA_{(p-1)\bbeta}=\bh_{y\bnu}\partial_{\bbeta}x^{\bnu}-\partial_{\bbeta}\ty \ ,
\nonumber \\
\end{eqnarray}
where $\bmu=0,1,\dots,7$ and we introduced dual coordinate $\ty=l_s^2 A_{p-1}$. Now we should distinguish two situations. In the first one we have $\tau_y\neq 0$. Then we find that the determinant $\det\bA_{\alpha\beta}$ has the form
\begin{eqnarray}
& &\det\bA_{\alpha\beta}=-\tau_y\tau_y
\det (\bh''_{\bmu\bnu}\partial_{\balpha}x^{\bmu}
\partial_{\bbeta}x^{\bnu}+\frac{\bh_{yy}}{\tau_y\tau_y}\partial_{\balpha}
\eta\partial_{\bbeta}\eta+ \nonumber \\
& &+B''_{\bmu\eta}\partial_{\balpha}x^{\bmu}\partial_{\bbeta}\eta+
B''_{\eta\bnu}\partial_{\balpha}\eta\partial_{\bbeta}x^{\bnu}+B''_{\bmu\ty}\partial_{\balpha}x^{\bmu}\partial_{\bbeta}\ty+
B''_{\ty\bnu}\partial_{\balpha}\ty\partial_{\bbeta}x^{\bnu}+l_s^2 F_{\balpha\bbeta}) \ , \nonumber \\
\end{eqnarray}
where
\begin{eqnarray}
& &\bh''_{\bmu\bnu}=
\bh_{\bmu\bnu}+\frac{\bh_{yy}}{\tau_y\tau_y}\tau_{\bmu}\tau_{\bnu}-
\frac{\tau_{\bmu}}{\tau_y}\bh_{y\bnu}-
\frac{\tau_{\bnu}}{\tau_y}\bh_{y\bmu} \ ,
\quad
\bh''_{\ty\ty}=\bh''_{\ty \bmu}=0 \ , \nonumber \\
& & B''_{\bmu\ty}=
-\frac{\tau_{\bmu}}{\tau_y} \ ,  \quad
B''_{\ty\bnu}=
\frac{\tau_{\bmu}}{\tau_y} \ , \nonumber \\
& & B''_{\bmu \eta}=
-(\tau_{\bmu}-\frac{\tau_y}{\bh_{yy}}\bh_{y\bmu})
\frac{\bh_{yy}}{\tau_y\tau_y} \ ,
\quad
B''_{\eta \bnu}=(\tau_{\bnu}-\frac{\tau_y}{\bh_{yy}}\bh_{y\bnu})
\frac{\bh_{yy}}{\tau_y\tau_y} \ . \nonumber \\
\end{eqnarray}
The resulting D(p-2)-brane is localized in $\eta$ and $\ty$ direction. However
since $h''_{\ty \bmu}=h''_{\ty\ty}=h''_{\ty \eta}=0$ there is no kinetic term for $\ty$ so that physical interpretation of this configuration is unclear.
For that reason we rather restrict ourselves to the case when $\tau_y=0$. This is natural
choice that implies that non-relativistic D(p-1)-brane wraps spatial direction only.
In this case we obtain
\begin{eqnarray}
\det \bA_{\alpha\beta}=
\bh_{yy}
\left|\begin{array}{cc}
\hbA'_{\balpha\bbeta} &
\tau_{\balpha}+\partial_{\balpha}\eta  \\
\tau_{\bbeta}-\partial_{\bbeta}\eta  & 0   \\
\end{array}\right|  \ ,
\nonumber \\
\end{eqnarray}
where
\begin{eqnarray}
\hbA'_{\balpha\bbeta}=\bh'_{\bmu\bnu}\partial_{\balpha}x^{\bmu}\partial_{\bbeta}x^{\bnu}+
\bh'_{\ty\ty}\partial_{\balpha}\ty\partial_{\bbeta}\ty+l_s^2 F_{\balpha\bbeta}+
B'_{\bmu \ty}\partial_{\balpha}x^{\bmu}\partial_{\bbeta}\ty+
B'_{\ty\bnu}\partial_{\balpha}\ty\partial_{\bbeta}x^{\bnu}+l_s^2 F_{\balpha\bbeta}
 \ , \nonumber \\
\end{eqnarray}
where
\begin{eqnarray}
& &\bh'_{\bmu\bnu}=
\bh_{\bmu\bnu}-\bh_{\bmu y}\frac{1}{\bh_{yy}}
\bh_{y\bnu} \ , \quad \bh'_{\ty\ty}=\frac{1}{\bh_{yy}} \ , \nonumber \\
& &B'_{\bmu \ty}=-\frac{\bh_{\bmu y}}{\bh_{yy}} \ , \quad
B'_{\ty \bnu}=\frac{\bh_{y\bnu}}{\bh_{yy}} \ . \nonumber \\
\end{eqnarray}
These are standard Buscher's rules
\cite{Buscher:1987sk,Buscher:1987sk}
 for components of T-dual metric in absence of NSNS two form.
As a result we obtain following action for non-relativistic D(p-2)-brane
\begin{equation}\label{nonrelp-2}
S=-T_{p-2}\int d^{p-2}\xi e^{-\phi'}
\sqrt{-\det\hbA_{\balpha\bbeta}}\sqrt{(\partial_{\balpha}\eta-\tau_{\balpha})
\hbA^{\balpha \bbeta}(\partial_{\bbeta}\eta+\tau_{\bbeta})} \ ,
\end{equation}
where
\begin{equation}
T_{p-2}=2\pi R_{y}T_{p-1} \ , \quad
\end{equation}
where $R_y$ is radius of the compact dimension $y$. Further, T-dual dilaton is equal to
\begin{equation}
\phi'=\phi-\frac{1}{2}\ln \sqrt{\bh_{yy}} \ .
 \end{equation}
In summary, we see that (\ref{nonrelp-2}) has manifestly the same form as
an action for non-relativistic D(p-1)-brane that shows that non-relativistic
D(p-1)-brane transforms covariantly under T-duality transformation  in the same
way as its relativistic precursor with however important condition that $\tau_y=0$.

\section{Non-Relativistic D-branes from Hamiltonian Formalism}\label{fifth}
\subsection{T-duality of D-brane in canonical formalism}
In this section we give an alternative definition of non-relativistic D-brane
that is based on the Hamiltonian formulation of Dp-brane. We begin with the
action for D-brane in the background with zero RR fields
\begin{eqnarray}
& &S=-T_p\int d^{p+1}\xi e^{-\phi}\sqrt{-\det \bA} \ ,
\nonumber \\
& & \bA_{\alpha\beta}=G_{MN}\partial_\alpha x^M\partial_\beta x^N+B_{MN}
\partial_\alpha x^M\partial_\beta x^N+l_s^2
F_{\alpha\beta} \ ,  \quad F_{\alpha\beta}=\partial_\alpha A_\beta-\partial_\beta A_\alpha \ .
\nonumber \\
\end{eqnarray}
From this action we obtain conjugate momenta
\begin{eqnarray}
& &p_M=-T_p e^{-\phi}(G_{MN}\partial_\beta x^N \bAi^{\beta 0}_S+
B_{MN}\partial_\beta x^N \bAi^{
\beta 0}_A)
\sqrt{-\det \bA} \ ,
\nonumber \\
& &\pi^i=-T_p l_s^2 e^{-\phi}\bAi^{i0}_A\sqrt{-\det\bA} \ .
\nonumber \\
\end{eqnarray}
Using these definitions we obtain following
 $p-$first class constraints
\begin{eqnarray}
\mH_i=p_M\partial_i x^M+F_{ij}\pi^j\approx 0\nonumber \\
\end{eqnarray}
together with following Hamiltonian constraint
\begin{eqnarray}
& &\mH_T\equiv \Pi_M G^{MN}\Pi_N+l_s^{-4}\pi^i \partial_i x^M G_{MN}\partial_j x^N\pi^j+
\nonumber \\
& &+
T_p^2 e^{-2\phi}\det (G_{MN}\partial_i x^M\partial_j x^N+B_{MN}\partial_i x^M
\partial_j x^N+l_s^2 F_{ij})
\approx 0 \ , \nonumber \\
\end{eqnarray}
where
\begin{equation}
\Pi_M=p_M-l_s^{-2}B_{MN}\partial_i x^N\pi^i \ .
\end{equation}
Let us now show how T-duality can be described in canonical formalism. As usual we should presume
an existence of  direction with isometry, say $x^9\equiv y$ and also that Dp-brane wraps this direction. Formally this can be achieved by fixing one spatial diffeomorphism constraint with the help of the gauge fixing function
\begin{equation}
\mG_p\equiv y-\xi^p\approx 0  \ ,
\end{equation}
where we further presume that all world-volume fields do not depend on $\xi^p$. Now it is easy
to see that $\mG_p$ has non-zero Poisson bracket with $\mH_p$ so that
they are second class constraints that can be explicitly solved. In fact, we can solve
$\mH_p=0$ for $p_y$ as
\begin{equation}
p_y=\partial_{\hi} A_p \pi^{\hi} \ , \quad  \hi=1,\dots,p-1 \ .
\end{equation}
Then we have
\begin{eqnarray}
& &\Pi_\mu
=p_\mu-l_s^{-2}
B_{\mu\nu}\partial_{\hi}x^{\nu} \pi^{\hi}-l_s^{-2}B_{\mu y}\pi^p \ , \quad
\mu=0,1,\dots,8 \ , \nonumber \\
& &\Pi_y=
\partial_{\hi}A_p\pi^{\hi}-
l_s^{-2}B_{y\mu}\partial_{\hi}x^{\mu}\pi^{\hi}
 \ . \nonumber \\
\end{eqnarray}
In order to identify components of the metric let us analyse  determinant $\det \bA_{ij}$
that is equal to
\begin{eqnarray}
& &\det \bA_{ij}=\det (\bA_{\hi \hj}-\bA_{\hi y}\frac{1}{\bA_{yy}}
\bA_{y \hj})\bA_{yy}=
\nonumber \\
& &=G_{yy}\det (\partial_{\hi}x^\mu(G_{\mu\nu}-\frac{1}{G_{yy}}
(G_{\mu y}G_{y\nu}+B_{\mu y}B_{y\nu}))\partial_{\hj}x^\nu+\nonumber \\
&&+
\partial_{\hi}x^\mu (B_{\mu\nu}-\frac{1}{G_{yy}}
(G_{\mu y}B_{y\nu}+B_{\mu y}G_{y\nu}))\partial_{\hj}x^\nu+\nonumber \\
\nonumber \\
& &+l_s^4\partial_{\hi}A_p \frac{1}{G_{yy}}\partial_{\hj}A_p+
l_s^2 \partial_{\hi}x^\mu \frac{G_{\mu y}}{G_{yy}}\partial_{\hj}A_p
-l_s^2 \partial_{\hi}A_p\frac{G_{y\nu}}{G_{yy}}\partial_{\hj}x^\nu+l_s^2 \partial_{\hi}x^\mu \frac{B_{\mu y}}{G_{yy}}\partial_{\hj}A_p-
l_s^2 \partial_{\hi}A_p \frac{B_{y\nu}}{G_{yy}}\partial_{\hj}x^\nu) \ .
\nonumber \\
\end{eqnarray}
First of all we find following transformation rule for dilaton $\phi$ as
\begin{equation}
\phi'= \phi-\frac{1}{2}\ln G_{yy} \ .
\end{equation}
Further, we see that it is natural to identify $A_p$ with dual $y'-$coordinate as
\begin{equation}
y'=l_s^2 A_p \
\end{equation}
and also we can identify components of T-dual metric and NSNS two form field
\begin{eqnarray}\label{Tdualback}
& & G'_{\mu\nu}=G_{\mu\nu}-\frac{1}{G_{yy}}
(G_{\mu y}G_{y\nu}+B_{\mu y}B_{y\nu}) \ ,
\quad
B'_{\mu\nu}=B_{\mu\nu}-\frac{1}{G_{yy}}
(G_{\mu y}B_{y\nu}+B_{\mu y}G_{y\nu}) \ ,
\nonumber \\
& & G'_{y'y'}=\frac{1}{G_{yy}} \ , \quad
G'_{\mu y'}=\frac{G_{\mu y}}{G_{yy}} \ , \quad
G'_{y'\mu}=-\frac{B_{y\mu}}{G_{yy}} \ ,
\quad
B'_{iy'}=\frac{G_{\mu y}}{G_{yy}} \ , \quad B'_{y'\nu}=
-\frac{G_{y\nu}}{G_{yy}} \nonumber \\
\end{eqnarray}
which are famous Buscher's rules \cite{Buscher:1987sk,Buscher:1987sk}. Further, since
we identified $A_p$ with T-dual coordinate $y'$ we identify $\pi^p$ with corresponding
conjugate momentum as
\begin{equation}
p_{y'}=l_s^2 \pi^p \
\end{equation}
so that we obtain Hamiltonian constraint for D(p-1)-brane in T-dual background
where the components of the dual metric and NSNS two forms are given in
(\ref{Tdualback})
\begin{eqnarray}
& &\mH'_T\equiv \Pi'_M G'^{MN}\Pi'_N+l_s^{-4}\pi^{\hi} \partial_{\hi} x'^M G'_{MN}\partial_{\hj} x'^N\pi^{\hj}+\nonumber \\
& &T_p^2 e^{-2\phi'}\det (G'_{MN}\partial_{\hi} x'^M\partial_{\hj} x'^N+B'_{MN}\partial_{\hi} x'^M
\partial_{\hj} x'^N+l_s^2 F_{\hi\hj}) \ .
\approx 0 \nonumber \\
\end{eqnarray}
However we should be more careful about scaling of various fields. Note that $\mH_T$ is world-sheet density of the dimension $[l_s^{-2p-2}]$. Since $H=\int d^p\xi N \mH_T$ should have dimension $l_s^{-1}$ we find that $N$ has dimension $[l_s^{1+p}]$. On the other hand T-dual Hamiltonian is given as
\begin{equation}
H'=\int dy \int d\xi^{p-1}N \mH'_T=
\int d \xi^{p-1} \left(\frac{N}{2\pi R}\right)
\mH'_T (2\pi R)^2 \ ,
\end{equation}
where $R$ is radius of compact dimension. We see that we should identify tension of T-dual brane as
\begin{equation}
T_{p-1}=2\pi R T_p
\end{equation}
and perform corresponding rescaling of conjugate momenta
$p_M'(2\pi R)\rightarrow p_M \ , \pi (2\pi R)\rightarrow \pi$ together with $N'=\frac{N}{2\pi R}$. Then  $\mH'_T$
and $N'$  have  correct dimensions.

\section{T-duality along Null Direction}\label{sixth}
In  previous section we demonstrated how T-duality can be described with the
help of canonical formalism. In this section we consider T-duality in
case of the background with null isometry given in (\ref{backmetin})
when the Hamiltonian constraint has the form
\begin{eqnarray}
& &\mH_\tau
=2p_u\Phi p_u-2p_u\hv^\mu p_\mu+p_\mu h^{\mu\nu}p_\nu+\nonumber \\
& &+2l_s^{-4}\pi^i\partial_i u \tau_\mu\partial_j x^\mu \pi^j+
l_s^{-4} \pi^i \partial_i x^\mu \bh_{\mu\nu}\partial_j x^\nu\pi^j+\nonumber \\
& &+T_p^2 e^{-2\phi}\det (\partial_i u\tau_\mu\partial_j x^\mu+
\partial_i x^\mu \tau_\mu\partial_j u+\partial_i x^\mu \bh_{\mu\nu}\partial_j x^\nu+
l_s^2 F_{ij}) \ .   \nonumber \\
\end{eqnarray}
Let us now presume that Dp-brane is extended in $u-$direction
which  means that we have gauge fixing function
$\mG=\xi^p-u \approx 0$ so that the diffeomorphism constraint $\mH_p$ can be solved for
$p_u$ as
\begin{equation}
p_u=\partial_{\hi} A_p \pi^{\hi} \
\end{equation}
while the matrix $\bA_{ij}$ has the form
\begin{equation}
\bA_{ij}=\left(\begin{array}{cc}
\bh_{\hi \hj}+l_s^2 F_{\hi \hj} & \partial_{\hi} x^\mu \tau_\mu+l_s^2 \partial_{\hi}A_p \\
\partial_{\hj}x^\mu \tau_\mu -l_s^2\partial_{\hj}A_p & 0 \\
\end{array}\right) \ .
\end{equation}
Let us presume that $\ba_{\hi \hj}=\bh_{\hi \hj}+l_s^2 \bF_{\hi \hj}$ is non-singular matrix with inverse $\ba^{\hi \hj}$ so that $\ba^{\hi \hj}\ba_{\hj \hk}=\delta^{\hi}_{\hk}$. Then we obtain
\begin{eqnarray}
\det\bA_{ij}=
\det \ba_{\hi \hj}(\partial_{\hi}\eta-\partial_{\hi}
x^\mu \tau_\mu) \ba^{\hi \hj}(\partial_{\hj}\eta+\partial_{\hj}
x^\nu \tau_\nu)
\nonumber \\
\end{eqnarray}
so that the Hamiltonian constraint has the form
\begin{eqnarray}
& &\mH_\tau'=2l_s^{-4}\partial_{\hi}\eta \pi^{\hi}
\Phi \partial_{\hj}\eta \pi^{\hj}-2l_s^{-2}\partial_{\hi}
\eta \pi^{\hi}\hv^\mu p_\mu+p_\mu h^{\mu\nu}p_\nu+
\nonumber \\
& &+2l_s^{-2}p_\eta \tau_\mu\partial_{\hi} x^\mu \pi^{\hi}+l_s^{-4}
\pi^{\hi}\partial_{\hi} x^\mu \bh_{\mu\nu}\partial_{\hj}x^\nu \pi^{\hj}+
\nonumber \\
& & +T_{(p-1)}^2 e^{-2\phi' }\det \ba_{\hi \hj}(\partial_{\hi}\eta-\partial_{\hi}
x^\mu \tau_\mu) \ba^{\hi \hj}(\partial_{\hj}\eta+\partial_{\hj}
x^\nu \tau_\nu) \ ,
\nonumber \\
\end{eqnarray}
where we identified $\eta=l_s^2A_p \ , p_\eta=l_s^{-2}\pi^p$ and we also introduced 
tension of D(p-1)-brane $T_{(p-1)}$ together with  dilaton $\phi'$ 
that are given by (\ref{defphi}).
We see that this Hamiltonian constraint coincides with the Hamiltonian constraint
derived in section (\ref{third}) which is again very nice consistency check.
Finally note that  spatial diffeomorphism constraints have the form
\begin{eqnarray}
\mH_{\hi}=
p_\mu \partial_{\hi}x^\mu+\partial_{\hi}\eta p_{\eta}+F_{\hi\hj}\pi^{\hj} \ .
\nonumber \\
\end{eqnarray}
\subsection{Special case: D1-brane}
It is interesting to consider Hamiltonian for the simplest example of non-relativistic
D-brane which is D1-brane. Since $\partial_\sigma \pi^\sigma\approx 0$ is the first class
constraint we can fix  it by imposing condition that $A_\sigma=\mathrm{const}$. Then
we can impose the condition that $\pi^\sigma=m$ so that the Hamiltonian constraint has the form
\begin{eqnarray}\label{mHD1}
& &\mH_\tau=2l_s^{-4}m^2\partial_\sigma \eta \Phi\partial_\sigma \eta-2ml_s^{-2}\partial_\sigma \eta
\hv^\mu p_\mu+p_\mu h^{\mu\nu}p_\nu+\nonumber \\
& &+2m l_s^{-2} p_\eta \tau_\mu \partial_\sigma x^\mu+m^2 l_s^{-4}\partial_\sigma x^\mu
\bh_{\mu\nu}\partial_\sigma x^\nu+T_p^2 e^{-2\phi}
(\partial_\sigma \eta \partial_\sigma \eta-\partial_\sigma x^\mu
\tau_\mu \partial_\sigma x^\nu\tau_\nu) \ .  \nonumber \\
\end{eqnarray}
Now the first part correspond to $m$ fundamental non-relativistic strings as can be easily seen comparing it with the Hamiltonian constraint for non-relativistic string which was found in \cite{Kluson:2018egd}. Note that it is crucial that $m$ is non-zero.

As the next step we determine Lagrangian density from (\ref{mHD1}). Using $H=\int d\sigma
(N^\tau \mH_\tau+N^\sigma \mH_\sigma)$ we obtain
\begin{eqnarray}
& &\partial_\tau x^\mu=\pb{x^\mu,H}=2N^\tau h^{\mu\nu}p_\nu-2N^\tau ml_s^{-2}\partial_\sigma \eta \hv^\mu+N^\sigma\partial_\sigma x^\mu \ ,
\nonumber \\
& &\partial_\tau \eta=\pb{\eta,H}=2N^\tau m l_s^{-2}\tau_\mu \partial_\sigma x^\mu+N^\sigma \partial_\sigma \eta \ .
\nonumber \\
\end{eqnarray}
Then we obtain  Lagrangian density in the form
\begin{eqnarray}
& &\mL=\frac{1}{4N^\tau}(\partial_\tau x^\mu \he_\mu^{ \ a}
\he_\nu^{ \ b}\partial_\tau x^\nu-2N^\sigma
\partial_\tau x^\mu \he_\mu^{ \ a}
\he_\nu^{ \ b}\partial_\sigma x^\nu
+(N^\sigma)^2
\partial_\sigma x^\mu \he_\mu^{ \ a}
\he_\nu^{ \ b}\partial_\sigma x^\nu)-\nonumber \\
& &-2N^\tau l_s^{-4}m^2\partial_\sigma \eta \Phi\partial_\sigma \eta
- N^\tau l_s^{-4} m^2\partial_\sigma x^\mu
\bh_{\mu\nu}\partial_\sigma x^\nu-N^\tau T_1^2 e^{-2\phi}
(\partial_\sigma \eta \partial_\sigma \eta-\partial_\sigma x^\mu
\tau_\mu \partial_\sigma x^\nu\tau_\nu) \ .
\nonumber \\
\end{eqnarray}
Finally we determine Lagrange multipliers $N^\tau,N^\sigma$. From the equation of  for $x^\mu$ and  $\eta$ we obtain
%
\begin{equation}
N^\sigma=\frac{\tau_{\tau}\tau_\sigma-\partial_\tau \eta\partial_\sigma \eta}{
\tau_\sigma \tau_\sigma-\partial_\sigma \eta \partial_\sigma \eta}
\ , \quad
N^\tau=\frac{\partial_\tau \eta\tau_\sigma-\tau_\tau\partial_\sigma\eta}
{2m l_s^{-2}(\tau_\sigma\tau_\sigma-
	\partial_\sigma \eta\partial_\sigma\eta)} \ .
\end{equation}
It can be shown as in \cite{Kluson:2018egd} that contribution proportional to $\Phi$ cancel
each other and hence Lagrangian density has the form
\begin{eqnarray}
& &\mL =\frac{m}{2l_s^2}
\frac{(\tau_\sigma\tau_\sigma-\partial_\sigma \eta\partial_\sigma \eta)}{
\partial_\tau \eta\tau_\sigma-\tau_\tau\partial_\sigma\eta}\bh_{\tau\tau}
-\frac{m}{l_s^2}\frac{(\tau_\tau\tau_\sigma-\partial_\tau \eta\partial_\sigma \eta)}
{\partial_\tau \eta\tau_\sigma-\tau_\tau \partial_\sigma\eta}
 h_{\tau\sigma}+\frac{m}{2l_s^2}\frac{(\tau_\tau\tau_\tau-\partial_\tau \eta\partial_\tau \eta)}
 {(\partial_\tau \eta\tau_\sigma-\tau_\tau \partial_\sigma \eta)}\bh_{\sigma\sigma}+
 \nonumber\\
 & &+\frac{1}{2ml_s^2}e^{-2\phi}(\partial_\tau\eta\tau_\sigma-\tau_\tau \partial_\sigma \eta)
\ . \nonumber \\
\end{eqnarray}
Again we see that the expression on the first line corresponds to $m$ coincided fundamental
non-relativistic strings while the expression on the last line has D-brane origin. We also see that there is crucial that $m\neq 0$ in order to find well defined Lagrangian density.

\end{document}